\documentclass[journal]{IEEEtran}

\ifCLASSINFOpdf
\else
   \usepackage[dvips]{graphicx}
\fi
\usepackage{url}

\usepackage{graphicx}
\usepackage{color}
\usepackage{cite}
\usepackage{amsmath}
\usepackage{amssymb}
\usepackage{subfigure}
\begin{document}

\title{Joint Design of Measurement Matrix and Sparse Support Recovery Method via Deep Auto-encoder}

%\author{\IEEEauthorblockN{A}
%\IEEEauthorblockA{A}
%}
\author{Shuaichao~Li,
        Wanqing~Zhang,
        Ying~Cui,
        Hei~Victor~Cheng,
        and~Wei~Yu% <-this % stops a space
\thanks{Manuscript received May 13, 2019; revised  July 17,  2019 and  August 30, 2019; accepted August 31, 2019.
This work was supported in part by NSFC China (61771309, 61671301, 61420106008, 61521062), and in part by NSERC Canada. The associate editor coordinating the review of this letter and approving it for publication was Prof. Wei Li. (Corresponding author: Ying Cui.)

S.~Li, W.~Zhang and Y. Cui are with Shanghai Jiao Tong University, Shanghai 200240, China (e-mail:
cuiying@sjtu.edu.cn). H. V. Cheng and W. Yu are with University of Toronto, Toronto, ON M5S3G4, Canada.
}}

\maketitle

\begin{abstract}
Sparse support recovery arises in many applications in communications and signal processing. Existing methods tackle sparse support recovery problems for a given measurement matrix, and cannot flexibly exploit the properties of sparsity patterns for improving performance. In this letter, we propose a data-driven approach to jointly design the measurement matrix and support recovery method for complex sparse signals, using auto-encoder in deep learning. The proposed architecture includes two components, an auto-encoder and a hard thresholding module. The proposed auto-encoder successfully handles complex signals using standard auto-encoder for real numbers. The proposed approach can effectively exploit properties of sparsity patterns, and is especially useful when these underlying properties do not have analytic models. In addition, the proposed approach can achieve sparse support recovery with low computational complexity. Experiments are conducted on an application example, device activity detection in grant-free massive access for massive machine type communications (mMTC). Numerical results show that the proposed approach achieves significantly better performance with much less computation time than classic methods, in the presence of extra structures in sparsity patterns.
\end{abstract}

\begin{IEEEkeywords}
Sparse support recovery, auto-encoder, deep learning, device activity detection, grant-free massive access
\end{IEEEkeywords}

\section{Introduction}

Sparse support recovery refers to the estimation of the locations of non-zero elements of a sparse signal of dimension $N$ based on a limited number of noisy linear measurements $L \ll N$. Sparse support recovery problems are of broad interest, with applications arising in various areas, such as subset selection in regression, structure estimation in graphical models, sparse approximation and signal denoising\cite{5319750}. There are two key challenges in sparse support recovery: designing a measurement matrix that can retain the information on sparsity while reducing the signal dimension, and recovering the support with low computational complexity based on the under-sampled linear measurements for a given measurement matrix. Existing works deal with these two challenges separately. Intuitively, jointly designing the measurement matrix and sparse support recovery method can maximally improve the performance of sparse support recovery. However, how to carry out such a joint design remains an open problem. In addition, existing works do not exploit structures of sparsity patterns. For many sparse support recovery problems in communications and signal processing, sparse signals have specific structures which may help improve the performance of sparse support recovery if properly used. However, many sparsity structures arising in practice do not have analytic models. In this letter, we address the aforementioned challenges using a \emph{data-driven} approach.

%For example, entries in the unknown signal vector are correlated with each other. Utilizing statistical dependence should improve the performance of the recovery. However, in most cases, the dependence is not known and hard to model. Moreover, even the dependence is known and given as prior information, adapting the sensing matrix to the structure is still an open problem.

%\textcolor{red}{
%Having these questions in mind, and motivated by the recent success of deep learning, we propose a data-driven approach which utilizes the deep auto-encoder structure to jointly design the sensing matrix and recovery module. The end-to-end training enables both the sensing matrix and the recovery module to learn and adapt to the underlying statistical properties from the data. With the presence of more statistical properties on the data, both the resulting measurement matrix and support recovery module are well adapt to sparse patterns. This provides a new promising approach to tackle the well-studied problem.}

Most existing works on sparse support recovery focus on tackling sparse support recovery problems for a given measurement matrix \cite{5319750, 5319742,4839045, 6994860,6158602, 5205769}. For example, in \cite{5319750, 5319742},  exhaustive methods are considered to derive conditions for exact sparse support recovery under the assumption that the sparsity level of the signal (i.e., the number of non-zero elements) is known in advance. The exhaustive methods have limited applications in practice due to their combinatorial complexity. In \cite{4839045, 6994860}, an optimization-based method, referred to as LASSO, is adopted for sparse support recovery with polynomial complexity  in $\mathcal{O}(N^3)$. In particular, \cite{4839045} directly deals with noisy linear measurements, while \cite{6994860} operates on the covariance matrix of noisy linear measurements. In \cite{6158602, 5205769}, assuming that the sparsity \cite{6158602} or the power order of the signal elements \cite{5205769} is known, the authors propose heuristic sparse support recovery algorithms which achieve lower computational complexity than LASSO at the cost of recovery performance loss. Note that none of \cite{5319750, 5319742, 4839045, 6994860, 6158602, 5205769} consider measurement matrix design, or exploit characteristics of sparsity patterns.

A closely related and more widely investigated topic is to estimate the sparse signal itself instead of its support. Most studies focus on recovering sparse signals under Gaussian measurement matrices which have certain performance guarantee \cite{tibshirani1996regression,8264818,8323218,yuan2006model}. Classic compressed sensing methods, such as LASSO \cite{tibshirani1996regression} with computational complexity of $\mathcal{O}(N^3)$ and AMP \cite{8264818,8323218} with (per iteration) computational complexity of $\mathcal{O}(LN)$, do not exploit hidden structures of the sparsity patterns. The performance of sparse signal recovery may be improved if additional properties of the sparsity patterns can be effectively exploited. For example, Group-LASSO \cite{yuan2006model} utilizes group sparsity which is assumed to be known \emph{a priori}. The authors in \cite{gregor2010learning,yao2017deepiot,7952561,taha2019enabling} exploit properties of sparsity patterns of real signals \cite{gregor2010learning,yao2017deepiot,7952561} and complex signals \cite{taha2019enabling} from training samples using data-driven approaches based on deep learning. To further improve performance, recent works \cite{sun2016deep,wu2019learning,mousavi2018a,nguyen2017deep,8262812,wu2019deep,8322184} consider joint design of signal compression and recovery methods using auto-encoder \cite{sun2016deep,wu2019learning,mousavi2018a,nguyen2017deep,8262812,8322184} and Generative Adversarial Networks (GAN) \cite{wu2019deep} in deep learning. In particular, \cite{sun2016deep,wu2019learning,mousavi2018a,nguyen2017deep} study linear compression for real signals; \cite{8262812,wu2019deep,8322184} consider nonlinear compression for real signals \cite{8262812,wu2019deep} and complex signals \cite{8322184}. Note that existing joint signal compression and recovery methods \cite{sun2016deep,wu2019learning,mousavi2018a,nguyen2017deep,8262812,wu2019deep,8322184} cannot provide linear compression for complex signals, and the extensions to joint linear compression and recovery methods for complex signals estimation are not trivial. In addition, the optimal measurement matrix and recovery method for sparse signal estimation are not necessarily always the best for support recovery.

%The main result of this paper shows that an auto-encoder approach can indeed be used to jointly design the linear measurement matrix and sparse support recovery method for complex signals. This paper further quantifies the performance improvement of this approach due to exploiting the structures of the sparsity patterns using numerical experiments.

%Measurement matrix designed for compressed sensing or compressed sensing methods can be used for sparse support recovery. However, one with better signal estimation does not necessarily provide better support recovery. It is not clear whether auto-encoder can be used to jointly design measurement matrix (corresponding to linear compression) and support recovery for complex signals, and  how much improvement for sparse support recovery can be achieved by exploiting properties on sparse patterns.

In this letter, we propose a data-driven auto-encoder architecture to jointly design the measurement matrix and support recovery method for complex sparse signals, using deep auto-encoder. The proposed architecture includes an auto-encoder and a hard thresholding module. The auto-encoder consists of an encoder which mimics the noisy linear measurement process, and a decoder which approximately performs sparse support recovery from the under-sampled linear measurements. The proposed auto-encoder successfully handles complex signals using standard auto-encoder for real numbers. The data-driven approach is especially useful when the underlying structures of sparsity patterns are hard to model, and can achieve sparse support recovery with low computational complexity due to the parallelizable neural network architecture. Experiments are conducted on an application example: device activity detection in grant-free massive access for massive machine type communications (mMTC). Numerical results show that the proposed approach achieves significantly better performance with much less computation time than classic methods, in the presence of additional properties of sparsity patterns. The substantial gains derive from the effective joint design that exploits these structures.
%\textcolor{red}{Specifically, our experimental results show that when the there exist strong correlation between the support, our proposed joint design significantly outperform all traditional approaches. The gains increase with the strength of the correlation, and the highest gain is achieved when the correlation is $1$, which is known as the group sparse case. However, we emphasize that the proposed approach is most useful when the underlying statistical properties are hard to model, and our auto-encoder based joint design can learn these properties from the data.}

\section{Support Recovery}
%In many practical problems of science and technology, one encounters the task of inferring quantities of interest from measured information. For instance, in signal and image processing, one would like to reconstruct a signal from measured data. When the information acquisition process is linear, the problem reduces to solving a linear system of equations. In mathematical terms, the observed data $\mathbf{y}\in C^L$ is connected to the signal $\mathbf{x}\in C^N$ of interest via
The support of sparse signal $\mathbf{x} \triangleq (x_n)_{n \in \mathcal{N}}\in \mathbb{C}^N$ is defined as the set of locations of non-zero elements of $\mathbf{x}$, denoted by ${\rm supp}(\mathbf{x}) \triangleq \{n\in\mathcal{N}|x_{n} \neq 0\}$, where $\mathcal{N} \triangleq \{1,\cdots,N\}$. We say $\mathbf{x}$ is sparse if the number of non-zero elements of $\mathbf{x}$ is much smaller than its total number of elements, i.e., $|{\rm supp}(\mathbf{x})| \ll N $. Consider $L \ll N$ noisy linear measurements $\mathbf{y}\in \mathbb{C}^L$ of $\mathbf{x}$:
\vspace*{-0.2cm}
\begin{align}
\mathbf{y} = \mathbf{A}\mathbf{x}+\mathbf{z}
\end{align}
where $\mathbf{z}\sim\mathcal{CN}(\mathbf{0},\sigma^2\mathbf{I}_L)$ is the additive white Gaussian noise (AWGN), $\mathbf{I}_L$ denotes the $L \times L$ identity matrix, and $\mathbf{A}\in \mathbb{C}^{L\times N}$ is the measurement matrix. Let $\boldsymbol{\alpha}\triangleq (\alpha_n)_{n \in \mathcal{N}}$, where $\alpha_n\triangleq \mathbb{I}[x_n \not=0]$ and $\mathbb{I}[\cdot]$ represents the indicator function. That is, ${\rm supp}(\mathbf{x})=\{n \in \mathcal{N}|\alpha_n=1\}$. The problem of support recovery is that of estimating ${\rm supp}(\mathbf{x})$ (or $\boldsymbol{\alpha}$) based on $\mathbf{y}$.

Sparse support recovery arises in several signal processing areas and has vast applications. As an important application example, we consider grant-free massive access, which is  recently proposed to support mMTC for IoT \cite{8264818}. Specifically, we consider a single cell with one single-antenna base station (BS) and $N$ single-antenna devices. Consider one coherent time slot. We use $\alpha_n \in \{0,1\}$ to denote the active state of device $n$, where $\alpha_n=1$ means that device $n$ accesses the channel, and $\alpha_n=0$ otherwise. The device-activity patterns for IoT traffic are typically sporadic. Thus, the number of active devices, denoted by $K \triangleq \sum_{n \in \mathcal{N}} \alpha_n$, is usually much smaller than $N$, i.e., $K \ll N$. We use $h_{n}\in \mathbb{C}$ to denote the complex channel between the BS and device $n$. We can view $\alpha_nh_n$ as $x_n$. Therefore, $\mathbf{x} \triangleq (x_n)_{n \in \mathcal{N}}$ is sparse, as $|{\rm supp}(\mathbf{x})|=K \ll N$. In grant-free massive access, there are two phases, i.e., the pilot transmission phase and the data transmission phase, and each device $n$ has a unique pilot sequence $\mathbf{a}_n \in \mathbb{C}^L$, where the pilot length $L \ll N$. In the pilot transmission phase, active devices synchronously send their pilot sequences to the BS. The received signal vector at the BS can be expressed as (1) with $\mathbf{A}=[\mathbf{a}_n]_{n \in \mathcal{N}} \in \mathbb{C}^{L\times N}$ representing the pilot matrix. Based on $\mathbf{y}$ and $\mathbf{A}$, the BS detects device activities $\boldsymbol{\alpha}$ \big(i.e., estimates ${\rm supp}(\mathbf{x})$\big). Once device activities are detected, the channels of the active devices can be estimated through the classic MMSE method. Thus, we shall focus on detecting device activities $\boldsymbol{\alpha}$, which is a sparse support recovery problem.
%additive white Gaussian noise, and the ${\rm diag}(s_{1},\cdots,s_{k})$ denotes a diagonal matrix with elements $(s_{1},\cdots,s_{k})$.The main goal of the BS in the pilot transmission phase is to detect user activities by recovering $\boldsymbol{\alpha}$ from the noisy observation $\mathbf{y}_{\rm pilot}$. If we use noisy linear measurements $\mathbf{y}$ to denote $\mathbf{y}_{\rm pilot}$, measurement matrix $\mathbf{A}$ to denote pilot matrix $\mathbf{A}^{\rm pilot}$, sparse signal $\mathbf{x}$ to denote ${\rm diag}(\boldsymbol{\alpha})\mathbf{h}$, and $\mathbf{n}$ to denote $\mathbf{z}$, then the activity detection problem in pilot transmission phase can be equivalent to the sparse support recovery problem described above.

%In this work, we mainly try to recover the support of $\mathbf{x}$, which is the set of indices corresponding to the nonzero entries and can be denoted by the sparse activation vector $\mathbf{a}\in \{0,1\}^N$.

\section{Proposed Approach}
\begin{figure}[t]
\vspace*{-0.1cm}
\begin{center}
 \includegraphics[width=3.4in]{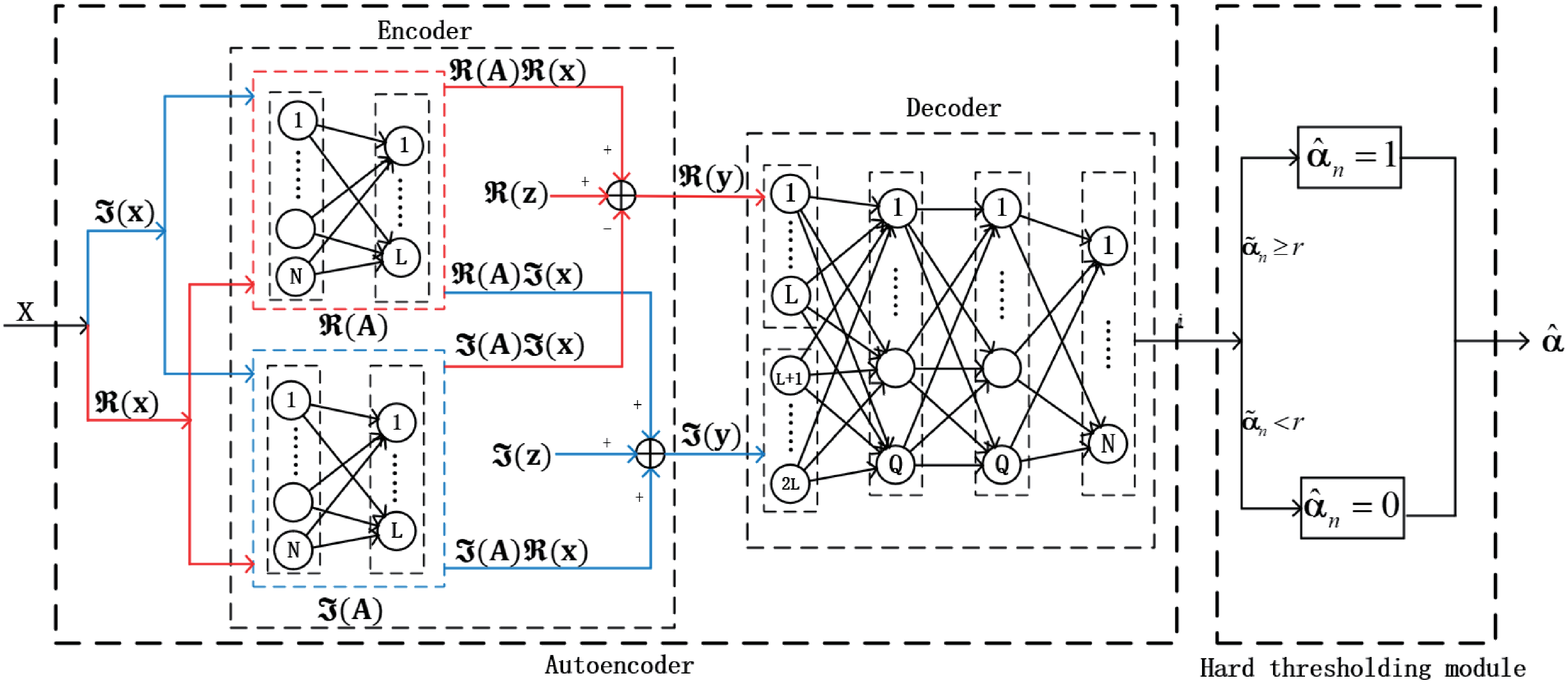}
  \end{center}
  \vspace*{-0.35cm}
     \caption{Proposed architecture.}
\label{networkmodel}
\vspace*{-0.55cm}
\end{figure}

In this section, we propose a data-driven approach to jointly design the measurement matrix and support recovery method for sparse complex signals. The proposed architecture consists of two components, an auto-encoder and a hard thresholding module, as illustrated in Fig.~\ref{networkmodel}.
%First, we introduce the converting module that converts the sparse signal $\mathbf{x}$ and measurement matrix $\mathbf{A}$ in terms of complex numbers to one in terms of real numbers. Then, we elaborate on the auto-encoder structure where the encoder mimics the noisy linear measurement process, and the encoder approximates the sparse support recovery process. Finally, we introduce the hard thresholding module which guarantees to achieve sparse support recovery.

\vspace*{-0.2cm}
\subsection{Auto-encoder}
Currently, standard neural networks can process only real numbers. However, sparse support recovery in many applications involves complex numbers. In this part, we introduce an auto-encoder for complex numbers using standard auto-encoder for real numbers in deep learning.
%with its encoder mimicking the noisy linear measurement process and decoder approximating the sparse support recovery process, using standard neural networks for real numbers.
%First we express the complex sensing matrix $A$ as $A = \Re(A) + j\Im(A)$, where $\Re(A),\Im(A)\in \mathbb{R}^{L\times N}$, and the complex vector $\mathbf{x},\mathbf{y}$ as $\mathbf{x}=\Re(\mathbf{x}) + j\Im(\mathbf{x}),\mathbf{y}=\Re(\mathbf{y}) + j\Im(\mathbf{y})$, where $\Re(\mathbf{y}),\Im(\mathbf{y})\in \mathbb{R}^L$ and $\Re(\mathbf{x}),\Im(\mathbf{x})\in \mathbb{R}^N$, then according to $\mathbf{y} = A\mathbf{x}+\mathbf{n}$, we can get the equations (2) and (3).
%\begin{align}
%\Re(\mathbf{y})=\Re(A)\Re(\mathbf{x})-\Im(A)\Im(\mathbf{x})+\Re(\mathbf{n})\\
%\Im(\mathbf{y})=\Im(A)\Re(\mathbf{x})+\Re(A)\Im(\mathbf{x})+\Im(\mathbf{n})
%\end{align}
%Therefore, the complex relation $\mathbf{y} = A\mathbf{x}+\mathbf{n}$ can be expressed equivalently via equations (2) and (3).

First, we introduce the encoder which mimics the noisy linear measurement process. The equation for complex numbers in (1) can be equivalently expressed via the following two equations for real numbers:
\vspace*{-0.1cm}
\begin{align}
\Re(\mathbf{y})=\Re(\mathbf{A})\Re(\mathbf{x})-\Im(\mathbf{A})\Im(\mathbf{x})+\Re(\mathbf{z})\\
\Im(\mathbf{y})=\Im(\mathbf{A})\Re(\mathbf{x})+\Re(\mathbf{A})\Im(\mathbf{x})+\Im(\mathbf{z})
\end{align}
where $\Re(\cdot)$ and $\Im(\cdot)$ represent the real part and imaginary part of a complex number.
Based on (2) and (3), we build two fully-connected neural networks, each with two layers, to mimic two linear relations with coefficient matrices $\Re(\mathbf{A})$ and $\Im(\mathbf{A})$, respectively. For each neural network, the input layer has $N$ neurons and the output layer has $L$ neurons; the weight of the connection from the $n$-th neuron in the input layer to the $l$-th neuron in the output layer represents the $(l,n)$-th element of the corresponding coefficient matrix; activation functions are not used in the output layer, to realize the linear relation. When $\Re(\mathbf{x})$ and $\Im(\mathbf{x})$ are input to the neural network corresponding to the linear relation with coefficient matrix $\Re(\mathbf{A})$ (or $\Im(\mathbf{A})$), $\Re(\mathbf{A})\Re(\mathbf{x})$ and $\Re(\mathbf{A})\Im(\mathbf{x})$ (or $\Im(\mathbf{A})\Re(\mathbf{x})$ and $\Im(\mathbf{A})\Im(\mathbf{x})$) can be obtained as the outputs. By summing $\Re(\mathbf{A})\Im(\mathbf{x})$, $\Im(\mathbf{A})\Re(\mathbf{x})$ and $\Im(\mathbf{z})$, $\Im(\mathbf{y})$ can be obtained using the encoder. By subtracting $\Im(\mathbf{A})\Im(\mathbf{x})$ from $\Re(\mathbf{A})\Re(\mathbf{x})$ and then adding $\Re(\mathbf{z})$, $\Re(\mathbf{y})$ can also be obtained using the encoder.

%\subsection{Autoencoder}

%\subsection{Decoder}
Next, we introduce the decoder which approximates the sparse support recovery process. We build a fully-connected neural network with four layers, i.e., one input layer, two hidden layers and one output layer. The input layer has $2L$ neurons with $\Re(\mathbf{y})$ being input into the first $L$ neurons and $\Im(\mathbf{y})$ being input into the last $L$ neurons. Each of the two hidden layers has $Q$ neurons and takes the rectified linear unit (ReLU), i.e., $ReLU(x)=\max(x,0)$, as the activation function. The output layer has $N$ neurons and uses Sigmoid function ($Sigmoid(x)=\frac{1}{1+e^{-x}}$) as the activation function to produce output $\tilde{\boldsymbol{\alpha}}\in [0,1]^N$ which is used to estimate $\boldsymbol{\alpha}$. We use matrixes $\boldsymbol{\Theta}_1\in \mathbb{R}^{Q\times 2L}$, $\boldsymbol{\Theta}_2\in \mathbb{R}^{Q\times Q}$ and $\boldsymbol{\Theta}_3\in \mathbb{R}^{N\times Q}$ to denote the weights of the connections from the input layer to the first hidden layer, the weights of the connections from the first hidden layer to the second hidden layer, and the weights of the connections from the second hidden layer to the output layer, respectively. We use vectors $\mathbf{b}_1\in \mathbb{R}^{Q} $, $\mathbf{b}_2\in \mathbb{R}^{Q} $ and $\mathbf{b}_3\in \mathbb{R}^{N} $ to denote the bias values corresponding to the first hidden layer, the second hidden layer and the output layer, respectively. We use $\mathbf{W} \triangleq ((\boldsymbol{\Theta}_i,\mathbf{b}_i))_{i=1,2,3}$ to denote the parameters of the decoder.%\subsection{Training the Auto-encoder}

We introduce the training procedure for the auto-encoder. Consider $I$ training samples $(\mathbf{x}^{(i)},\boldsymbol{\alpha}^{(i)}),i=1,\cdots,I$. We use $\tilde{\boldsymbol{\alpha}}^{(i)}$ to denote the output of the auto-encoder corresponding to input $\mathbf{x}^{(i)}$, which depends on $(\mathbf{W},\mathbf{A})$. To measure the distance between $\boldsymbol{\alpha}^{(i)}$ and $\tilde{\boldsymbol{\alpha}}^{(i)}$, we use the cross-entropy loss function:
\begin{small}
$L(\mathbf{W},\mathbf{A})=\frac{1}{NI}\sum_{i=1}^{I}\sum_{n=1}^{N}-(\alpha_n^{(i)}\log(\tilde{\alpha}_n^{(i)})+(1-\alpha_n^{(i)})\log(1-\tilde{\alpha}_n^{(i)})).$
\end{small}We train the auto-encoder using the ADAM algorithm which is a first-order gradient-based optimization algorithm for stochastic objective functions \cite{kingma2015adam}. After training, we obtain the measurement matrix by extracting the weights of the encoder. In addition, we can directly use the decoder to perform sparse support recovery.
\vspace*{-0.2cm}
\subsection{Hard Thresholding Module}
Note that even after training, there is no guarantee that the proposed architecture can produce an output $\tilde{\boldsymbol{\alpha}}$ that is in $\{0,1\}^N$. Thus, we build a hard thresholding module parameterized by threshold $r$ to convert the output of the auto-encoder $\tilde{\boldsymbol{\alpha}} \in [0,1]^N$ to the final output of the proposed architecture $\hat{\boldsymbol{\alpha}} \triangleq (\hat{\alpha}_n)_{n \in \mathcal{N}} \in \{0,1\}^N$, where $\hat{\alpha}_n \triangleq \mathbb{I}[\tilde{\alpha}_n\geq r]$. Let $P_E(r) \triangleq \frac{1}{I} \sum_{i=1}^I\frac{\|\boldsymbol{\alpha}^{(i)}-\hat{\boldsymbol{\alpha}}^{(i)} \|_1}{N}$ denote the error rate for a given threshold $r$. Given $I$ training samples $(\mathbf{x}^{(i)},\boldsymbol{\alpha}^{(i)}),i=1,\cdots,I$, we choose  the optimal threshold $r^*=\mathop{\arg\min}_{r}P_E(r)$, and use the optimized error rate $P_E^*=P_E(r^*)$ as the performance metric for the proposed approach.

\section{Numerical Results}
%In this section, we use $(x^{(i)},a^{(i)})$  to indicate the $i$th training sample and study three types of user activation patterns: the first is that the users are independently activated with the same probability, that is, the elements of the activation vector $\mathbf{a}$ are independently and identically distributed (This case will be referred to as Case 1 below). The purpose is to test the performance of the neural network under general conditions and use it as a benchmark. Second, the users are activated independently with different probabilities, that is, the elements in the sparse activation vector $\mathbf{a}$ are independently and non-identically distributed.  (This case will be referred to as Case 2 below). The purpose is to test the adaptability of the neural network to complex situations; the third is that the activation of the user is relevant, that is, the elements in the activation vector a are block activated (This case will be referred to as Case 3 below). The purpose is to see if the neural network can make good use of the a priori information contained in the sample.
In the simulation, we consider device activity detection in grant-free massive access with $N$ single-antenna devices each with a pilot sequence of length $L$ and one single-antenna BS. We show the average error rate of the proposed data-driven approach and five baseline schemes\footnote{The deep learning based methods in \cite{gregor2010learning,yao2017deepiot,7952561,sun2016deep,wu2019learning,mousavi2018a,nguyen2017deep,8262812,wu2019deep,8322184} are not applicable in our setup.} over the same set of testing samples. Specifically, we consider classic methods, i.e., LASSO with the optimal regularization parameter \cite{tibshirani1996regression}, Group LASSO \cite{yuan2006model}, Sparse Group LASSO \cite{simon2013sparse} and AMP \cite{8323218}. In addition, to demonstrate the effectiveness of the measurement matrix design via the encoder in the proposed architecture, we also consider a deep learning (DL) method, relying on the same structure as the decoder and hard thresholding module in the proposed architecture but without measurement matrix design \cite{taha2019enabling}.\footnote{Note that the obtained measurement matrix is in general not necessarily suitable for other sparse support recovery methods. The proposed measurement matrix design using auto-encoder is applicable to cases where a sparse support recovery method can be approximated using a neural network.} All baseline schemes adopt the same set of pilot sequences for the $N$ devices whose entries are independently generated from $\mathcal{CN}(0,1)$. To guarantee that the power of each pilot is the same as that for the baseline schemes, we require ${\|\mathbf{a}_n\|}_2=\sqrt{L}$ in training the proposed architecture.

In the simulation, we choose $h_n \sim \mathcal{CN}(0,1)$ and $\sigma^2=0.1$. To show how the proposed data-driven approach benefits from exploiting the structures of the sparsity patterns, we consider three cases. In Case 1, $N=40$ devices randomly access the channel in an i.i.d. manner with access probability $\Pr[\alpha_n=1]=p$, $n \in \mathcal{N}$. In Case 2, $N=40$ devices are divided into two groups, i.e., $\mathcal{N}_1$ and $\mathcal{N}_2$ of the same size with the devices in $\mathcal{N}_i$ accessing the channel in an i.i.d. manner with access probability $\Pr[\alpha_n=1]=p_i$, $n \in \mathcal{N}_i, i=1,2$, and let $p = \frac{p_1+p_2}{2}$. In Case 3, $N=200$ devices are divided into 40 groups of the same size, and 40 Bernoulli random variables $\xi_j, j \in \{1, \cdots, 40\}$ are i.i.d. with $\Pr[\xi_j=1]=p_g$, for all $j \in \{1, \cdots, 40\}$; if $\xi_j = 1$, all devices in the $j$-th group access the channel in an i.i.d. manner with access probability $p_u \in [0,1]$. Thus, $p_u$ can be treated as the conditional access probability for each device, and $p = p_up_g$ represents the access probability for each device. Note that the active states of the devices in one group are correlated for all $p_u \in [0,1]$, and are the same when $p_u =1$. Table~\ref{parameter} shows the sizes of training samples and validation samples for training the architectures in the proposed approach and the DL method, and the sizes of testing samples for evaluating the proposed approach and the baseline schemes. All testing samples are excluded from the training and validation samples. The maximization epochs, learning rate and batch size in training the proposed architecture are set as 100000, 0.001 and 128, respectively. When the value of the loss function on the validation set does not change for five epoches, the training process is stopped and the corresponding parameters of the auto-encoder are saved.

\begin{table}[t]
\centering
\caption{Sample sizes}
   \vspace{-0.1cm}
\begin{tabular}{|c|c|c|c|}
\hline
&Training&Validation&Testing\\
\hline
Case 1, Case 2&$4.5 \times 10^5$&$5\times10^4$&$1\times10^4$\\
\hline
Case 3&$9\times10^4$&$1\times10^4$&$1 \times10^5$\\
\hline
\end{tabular}
\label{parameter}
   \vspace{-0.6cm}
\end{table}

\begin{figure}[t]
\begin{center}
 \vspace*{-0.1cm}
 \subfigure[\scriptsize{Error rate versus $L/N$ at $p=0.1$. }]
 {\resizebox{4.2cm}{!}{\includegraphics{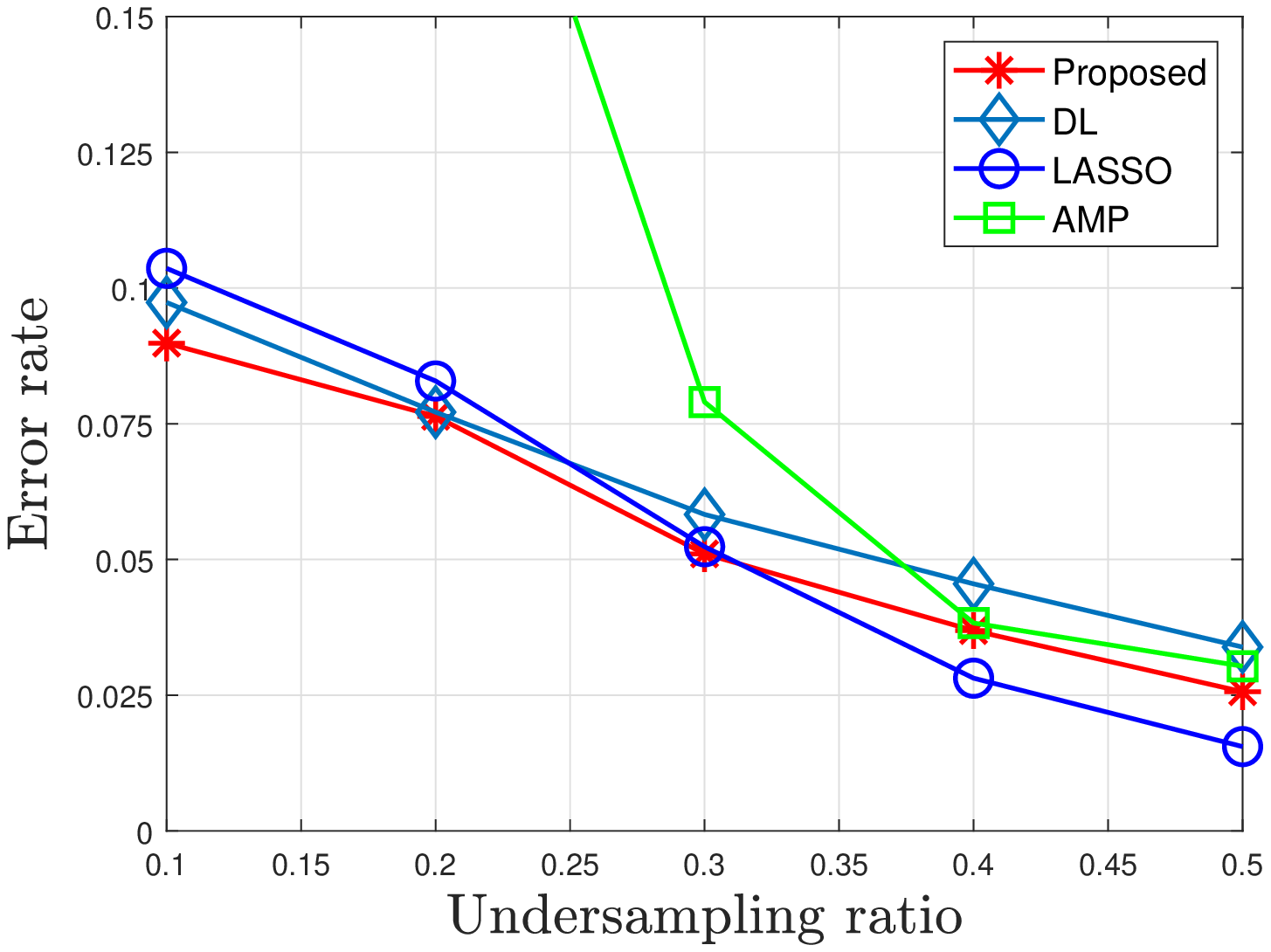}}}
 \subfigure[\scriptsize{Error rate versus $p$ at $L/N=0.3$.}]
 {\resizebox{4.2cm}{!}{\includegraphics{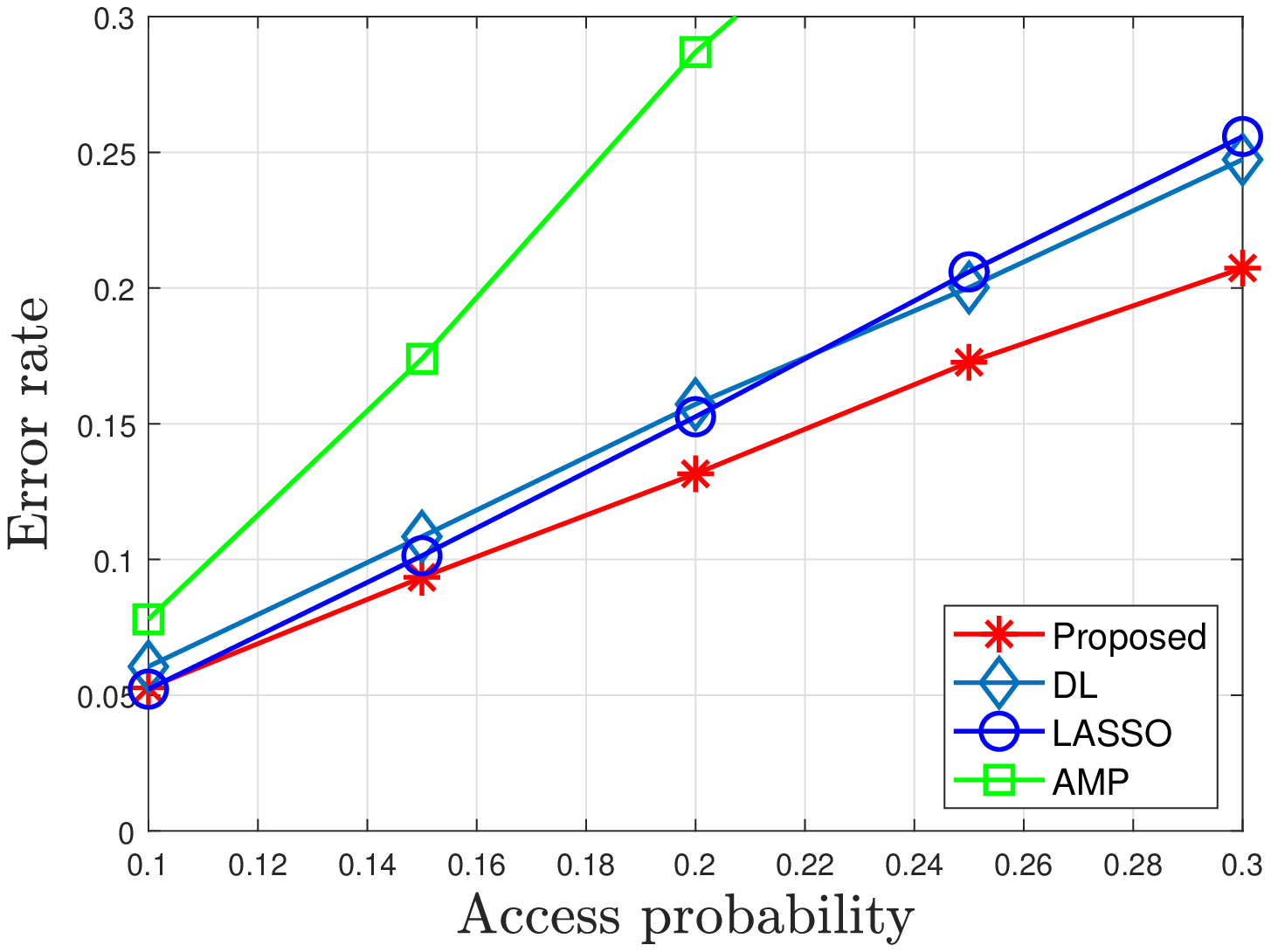}}}
 \end{center}
  \vspace*{-0.5cm}
   \caption{\small{Error rate versus undersampling ratio ($L/N$) and access probability ($p$) in Case 1.}}
   \label{case1}
\end{figure}

\begin{figure}[t]
\begin{center}
 \vspace*{-0.3cm}
 \subfigure[\scriptsize{Error rate versus $L/N$ at $p=0.1$ and $p_1/p_2=4$.}]
 {\resizebox{4.2cm}{!}{\includegraphics{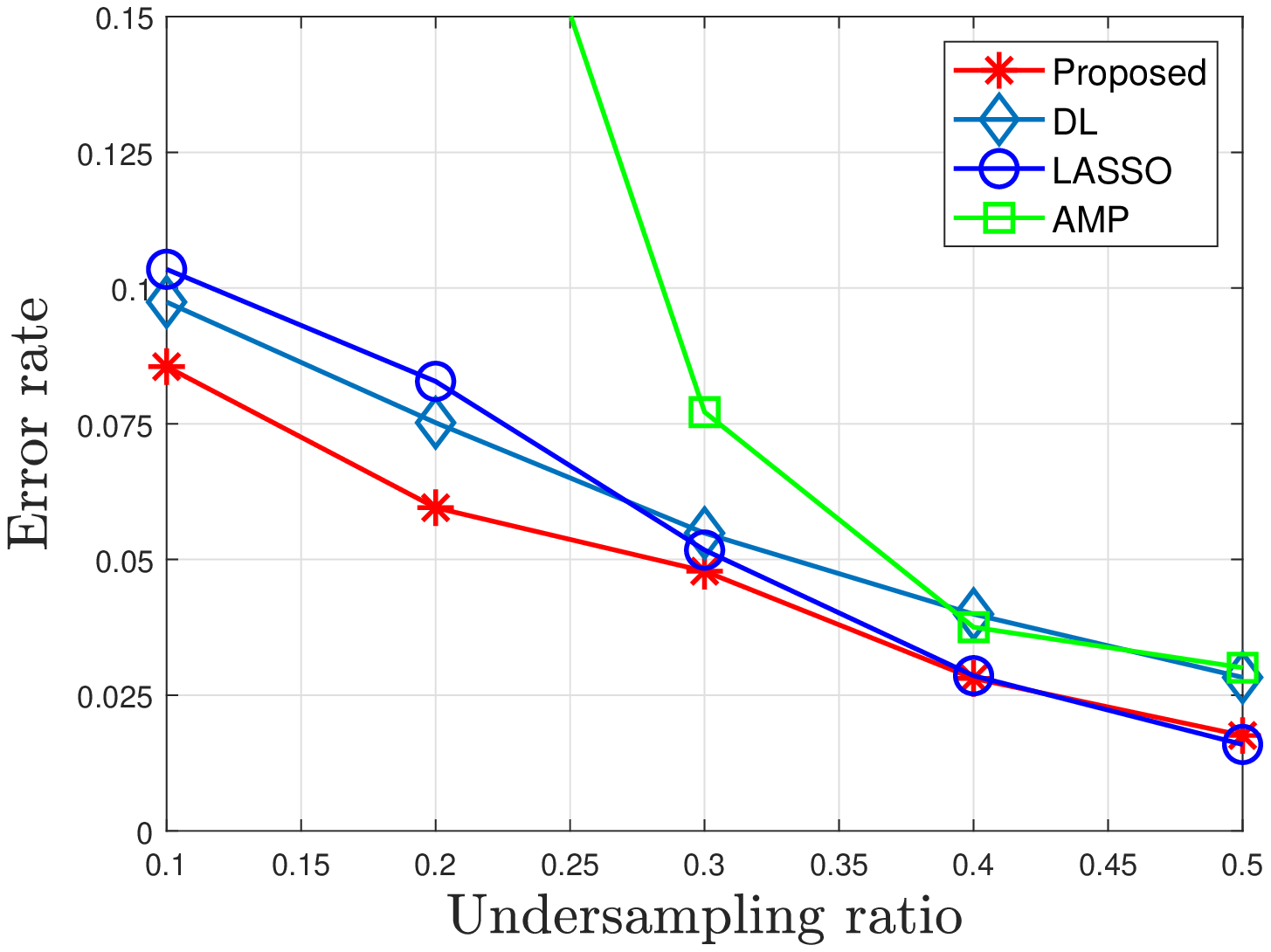}}}
 \subfigure[\scriptsize{Error rate versus $p$ at $L/N=0.3$ and $p_1/p_2=4$.}]
 {\resizebox{4.2cm}{!}{\includegraphics{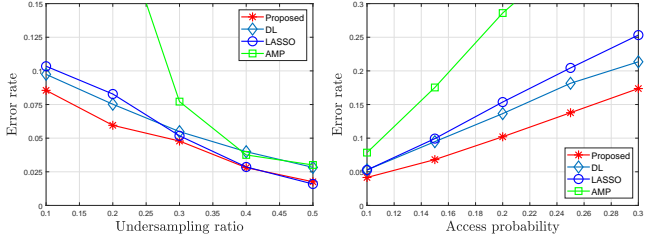}}}
  \subfigure[\scriptsize{Error rate versus $p_1/p_2$ at $p=0.1$, $L/N=0.3$.}]
 {\resizebox{4.2cm}{!}{\includegraphics{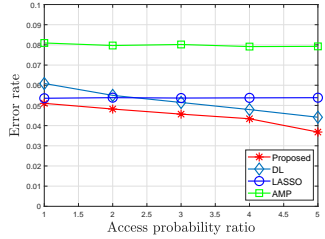}}}
 \end{center}
 \vspace*{-0.3cm}
   \caption{\small{Error rate versus undersampling ratio ($L/N$), access probability ($p$) and access probability ratio ($p_1/p_2$) in Case 2.}}
   \label{case2}
   \vspace{-0.2cm}
\end{figure}

\begin{figure}[t]
\begin{center}
\vspace*{-0.1cm}
 \subfigure[\scriptsize{Error rate versus $L/N$ at $p=0.1$, $p_u = 1$.}]
 {\resizebox{4.2cm}{!}{\includegraphics{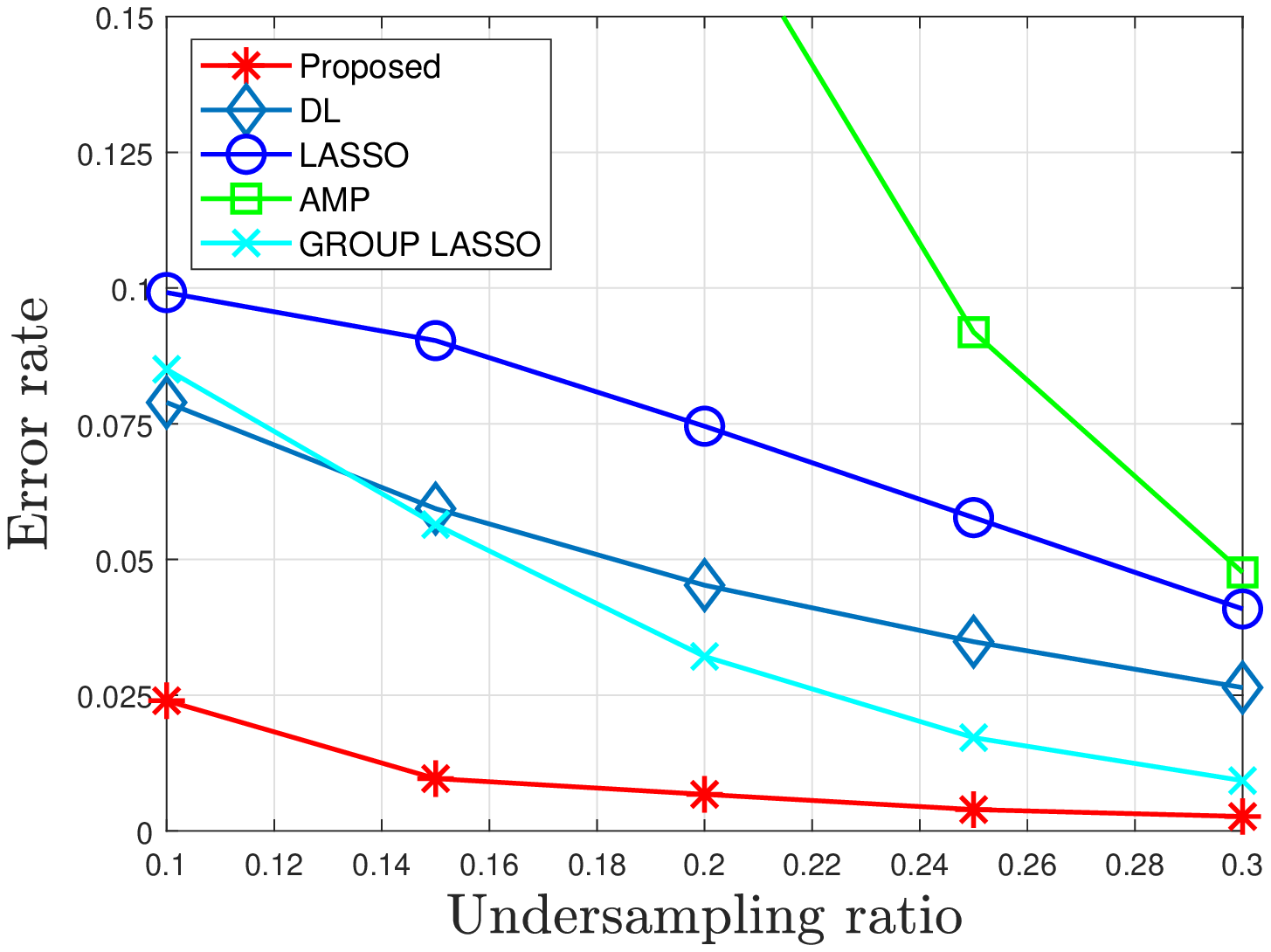}}}
   \subfigure[\scriptsize{Error rate versus $L/N$ at $p=0.1$, $p_u = 0.9$.}]
 {\resizebox{4.2cm}{!}{\includegraphics{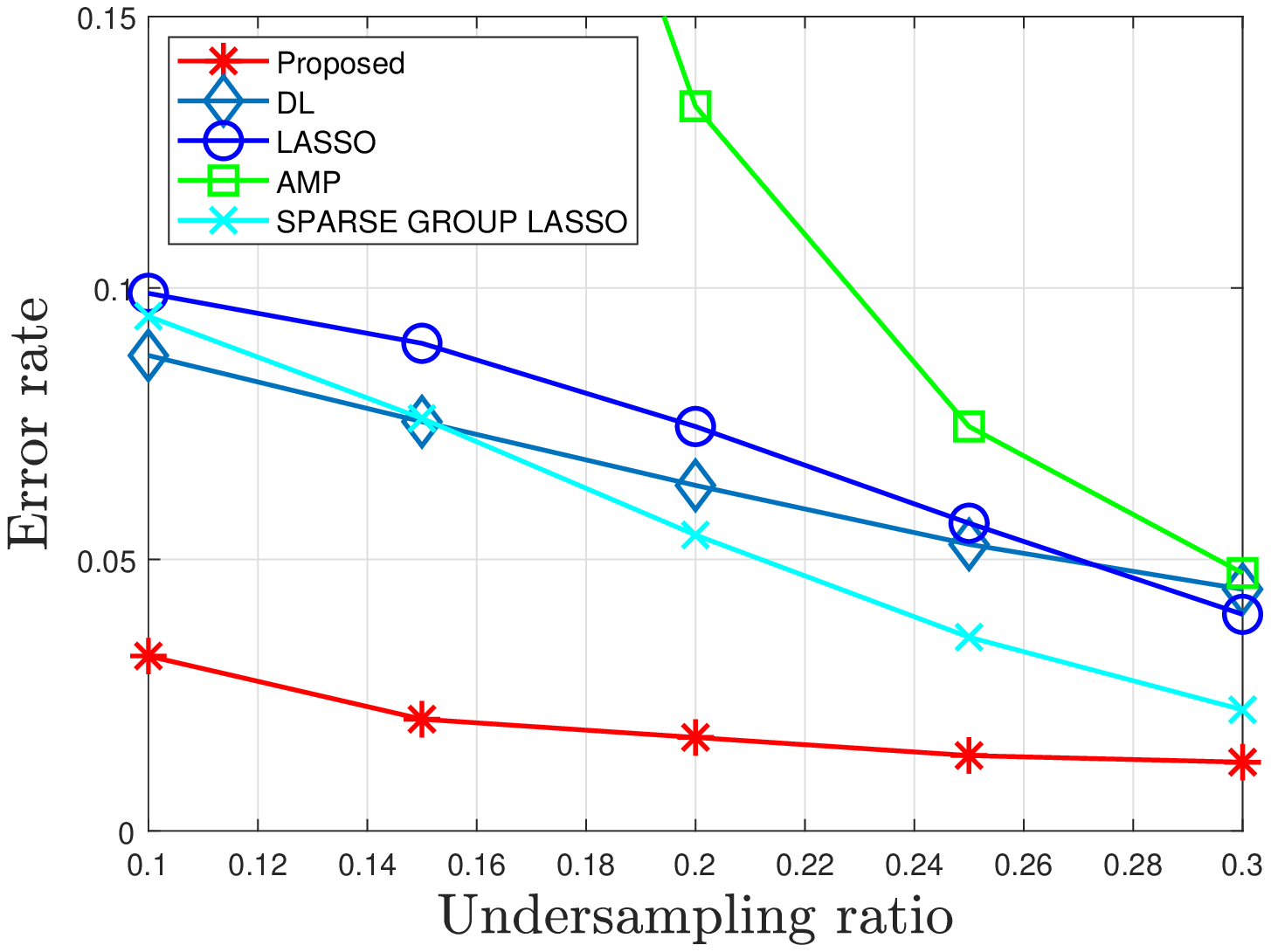}}}
 \subfigure[\scriptsize{Error rate versus $p$ at $L/N=0.3$, $p_u = 1$.}]
 {\resizebox{4.2cm}{!}{\includegraphics{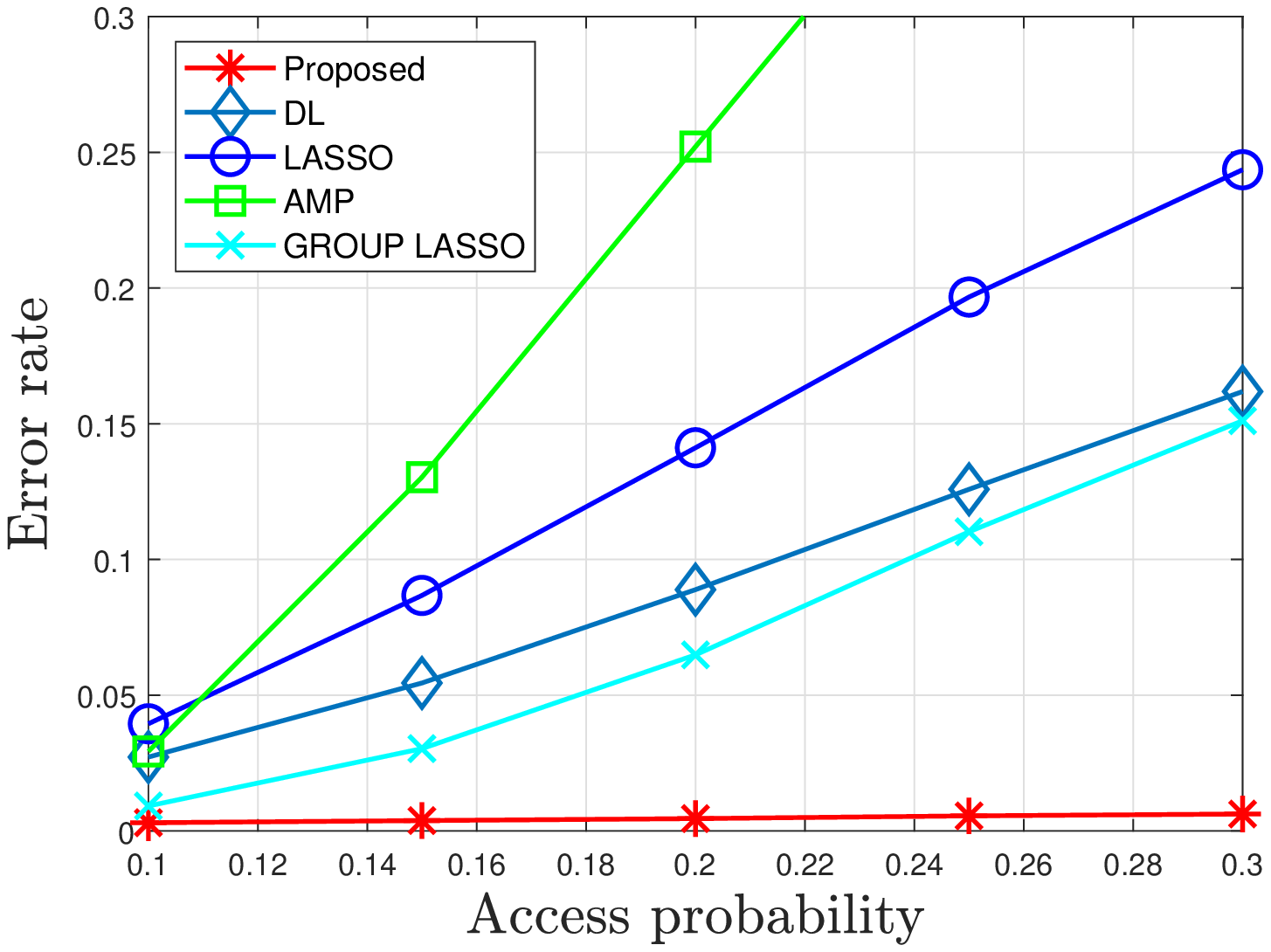}}}
 \subfigure[\scriptsize{Error rate versus $p_u$ at $L/N=0.3$, $p=0.1$.}]
 {\resizebox{4.2cm}{!}{\includegraphics{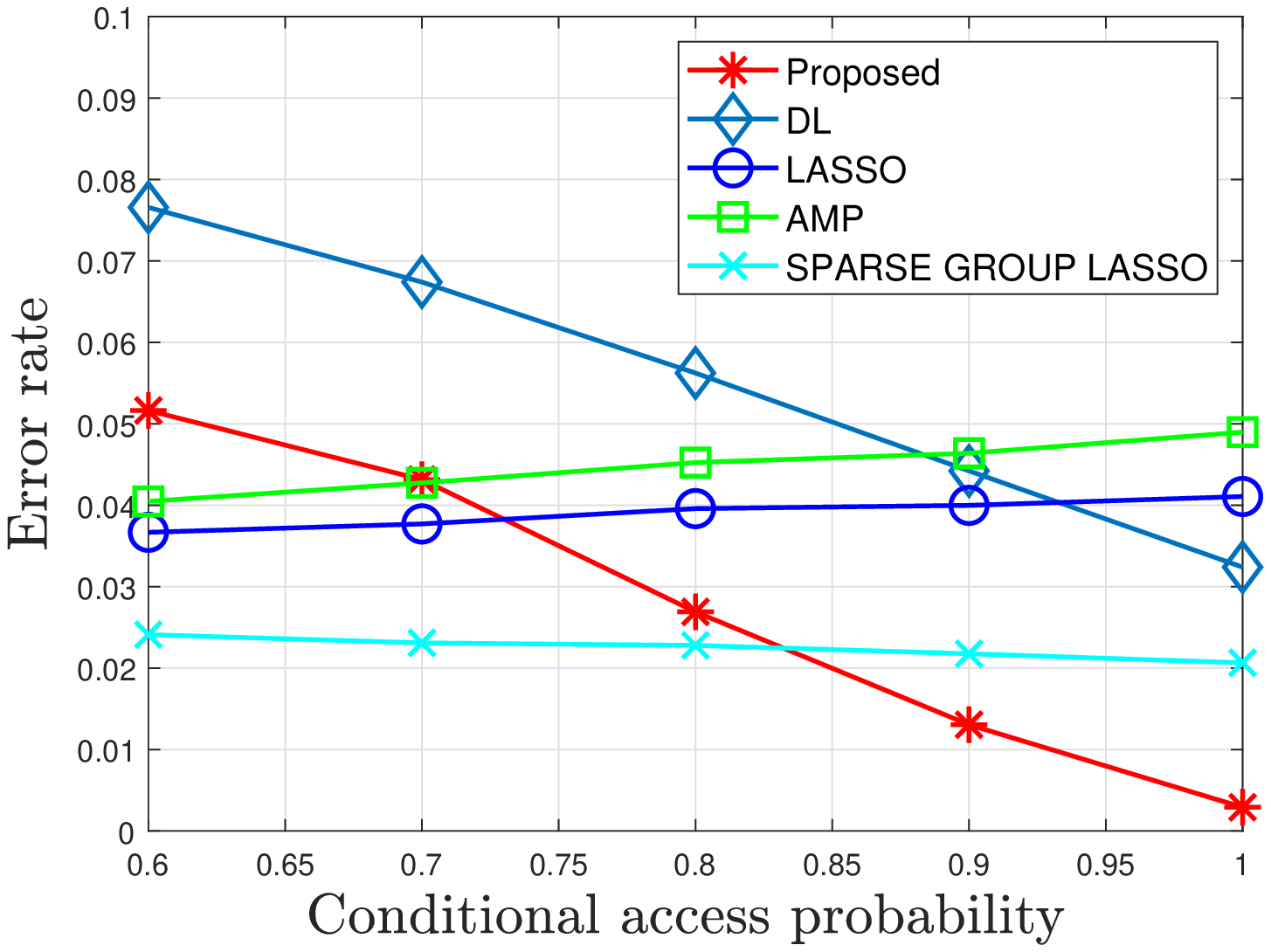}}}
  \end{center}
 \vspace*{-0.3cm}
   \caption{\small{Error rate versus undersampling ratio ($L/N$), access probability ($p$) and conditional access probability ($p_u$) in  Case 3.}}
   \label{case3}
   \vspace{-0.5cm}
\end{figure}

Fig.~\ref{case1}(a), Fig.~\ref{case2}(a) and Fig.~\ref{case3}(a),(b) illustrate the error rate of device activity detection versus the undersampling rate $L/N$ at $p = 0.1$ in Case 1, Case 2 and Case 3, respectively. Fig.~\ref{case1}(b), Fig.~\ref{case2}(b) and Fig.~\ref{case3}(c) illustrate the error rate of device activity detection versus the access probability $p$ at $L/N = 0.3$ in Case 1, Case 2 and Case 3, respectively. In each case, the error rate of each scheme decreases with $L/N$ and increases with $p$; LASSO outperforms AMP when $L/N$ is small or when $p$ is large, owing to its optimization framework, and performs similarly to AMP in the other regimes; the proposed data-driven approach outperforms DL, demonstrating the importance of measurement matrix design in improving sparse support recovery. In Case 1, the proposed data-driven approach has similar performance as LASSO, because no extra properties of the sparsity patterns can be extracted from the training samples to improve sparse support recovery. In Case 2, the gain of the proposed data-driven approach over LASSO is larger than that in Case 1, and the gain increases with $p_1/p_2$ as shown in Fig.~\ref{case2}(c), which shows that the proposed data-driven approach can exploit the difference in device activity for the two groups to improve sparse support recovery. In Case 3 with $p_u = 1$, we additionally consider Group LASSO \cite{yuan2006model}, which is specifically designed for group-wise sparse signals and explicitly utilizes the values of the group size and the number of groups. In Case 3 with $p_u \in (0,1)$, we additionally consider Sparse Group LASSO \cite{simon2013sparse}, which is specifically designed for group-wise and within group sparse signals and explicitly utilizes the values of the group size and the number of groups. By making use of group sparsity, Group LASSO and Sparse Group LASSO significantly outperform LASSO. In Case 3 with large $p_u$, the proposed data-driven approach overwhelmingly outperforms LASSO and even significantly outperforms Group-LASSO and Sparse Group LASSO, as it successfully exploits the properties of the sparsity patterns based on the training samples in designing both the measurement matrix and support recovery method; the performance of the proposed data-driven approach increases with $p_u$, while the performance of LASSO and Sparse Group LASSO remains almost the same, as shown in Fig.~\ref{case3}(d). Table~\ref{runtime} shows that the proposed sparse support recovery method (corresponding to the decoder and hard thresholding module in the proposed architecture implemented in tensorflow) runs much faster than the classic methods in all three cases.\footnote{LASSO and AMP are conducted using MATLAB, while the proposed method is implemented using Python. The corresponding running times are evaluated on the same server.}
\vspace{-0.2cm}
\begin{table}[!htbp]
\centering
\caption{CPU Running times (sec) for one sample at $p=0.1$ and $L/N=0.3$}
   \vspace{-0.4cm}
\begin{tabular}{|c|c|c|c|}
\hline
&Proposed method&LASSO&AMP\\
\hline
Cases 1,2&$1.1\times 10^{-5}$&$1.3 \times10^{-2}$&$8.4 \times10^{-2}$\\
\hline
Case 3&$1.3 \times10^{-5}$&$1.2 \times10^{-1}$&$3.9 \times10^{-1}$\\
\hline
\end{tabular}
\label{runtime}
   \vspace{-0.5cm}
\end{table}

\section{Conclusion}
In this letter, we propose a data-driven approach to jointly design the measurement matrix and support recovery method for complex sparse signals using the concept of auto-encoder in deep learning. Due to the effective joint design and the ability to exploit the structures of sparsity patterns, the proposed approach achieves a much lower error rate than classic methods. Furthermore, the proposed approach is able to achieve sparse recovery with high computational efficiency.

\bibliographystyle{IEEEtran}

\end{document}